

\input harvmac.tex

\def\pslash{p\!\!/}

\def\p2inf{\mathrel{\mathop{\sim}\limits_{\scriptscriptstyle
{p^2 \to \infty }}}}
\def\q2inf{\mathrel{\mathop{\sim}\limits_{\scriptscriptstyle
{q^2 \to \infty }}}}
\def\Q2inf{\mathrel{\mathop{\sim}\limits_{\scriptscriptstyle
{Q^2 \to \infty }}}}
\def\kap2inf{\mathrel{\mathop{\sim}\limits_{\scriptscriptstyle
{\kappa \to \infty }}}}
\def\x2inf{\mathrel{\mathop{\sim}\limits_{\scriptscriptstyle
{x \to \infty }}}}
\def\Lam2inf{\mathrel{\mathop{\sim}\limits_{\scriptscriptstyle
{\Lambda \to \infty }}}}

\def\frac#1#2{{{#1}\over {#2}}}
\def\half{\hbox{${1\over 2}$}}

\def\smallfrac#1#2{\hbox{${#1\over #2}$}}

\def\tr{{\rm tr}}
\def\Tr{{\rm Tr}}

\def\matele#1#2#3{\langle {#1} \vert {#2} \vert {#3} \rangle }
\def\PR{{\it Phys.~Rev.~}}
\def\PRL{{\it Phys.~Rev.~Lett.~}}
\def\NP{{\it Nucl.~Phys.~}}
\def\AP{{\it Annal.~Phys.~}}
\def\PRep{{\it Phys.~Rep.~}}
\def\PL{{\it Phys.~Lett.~}}
\def\ZP{{\it Zeit.~Phys.~}}
\def\NC{{\it Nuov.~Cim.~}}

\def\vyp#1#2#3{\hbox{{\bf #1} (#2) #3}}
\catcode`@=11 
\def\slash#1{\mathord{\mathpalette\c@ncel#1}}
 \def\c@ncel#1#2{\ooalign{$\hfil#1\mkern1mu/\hfil$\crcr$#1#2$}}
\def\lsim{\mathrel{\mathpalette\@versim<}}
\def\gsim{\mathrel{\mathpalette\@versim>}}
 \def\@versim#1#2{\lower0.2ex\vbox{\baselineskip\z@skip\lineskip\z@skip
       \lineskiplimit\z@\ialign{$\m@th#1\hfil##$\crcr#2\crcr\sim\crcr}}}
\catcode`@=12 
\def\Fig#1#2#3{\nfig\#1{#2}}

\pageno=0
\tolerance=10000
\hfuzz=5pt
\baselineskip 12pt plus 2pt minus 2pt
\nopagenumbers
\line{\hfil   Saclay-T93/138}
\line{\hfil December 1993}
\medskip
\centerline{\bf The Regularization of the Fermion Determinant}
\centerline{\bf in Chiral Quark Models}
\vskip 36pt
\centerline{Richard D.~Ball
}
\vskip 12pt
\centerline{\it Theory Division, CERN,}
\centerline{\it CH-1211 Gen\`eve 23, Switzerland.}
\vskip 12pt
\centerline{and}
\vskip 12pt
\centerline{Georges Ripka}
\vskip 12pt
\centerline{\it Service de Physique Theorique,}
\centerline{\it Centre d'Etudes de Saclay,}
\centerline{\it F-91191 Gif-sur-Yvette, France.}
\vskip 1.in
{\narrower\baselineskip 10pt
\centerline{\bf Abstract}
\medskip
The momentum dependence of the quark self energy gives a physically
motivated and consistent regularization of both the real and imaginary
parts of the quark loop contribution to the meson action. We show that
the amplitudes for anomalous processes are always reproduced correctly.
\vskip 10pt
}

\vskip .7in
\centerline{To be published in : {\it Proceedings of the Conference on Many
Body Physics,}}
\centerline {{\it Coimbra, Portugal, September 20-25, 1993.}}
\Date{}
\vfill
\eject
\noblackbox

The regularization of the fermion determinant in effective low energy
chiral quark models has been the subject of much debate, frustration and
confusion. The contribution of the fermion determinant to the meson
effective action
takes the form of a quark loop (see fig.~1) which
induces the propagation and interactions of the local chiral field.
If the couplings are local the loop integration is
formally infinite, and the model is then only
well defined when supplemented with some regularization and
renormalization prescription. The standard procedure works only
for a very limited set of models,
such as the linear sigma--model for example, so in general it is
necessary to introduce a finite cut--off scale and treat the meson
effective action as an effective field theory. In
such cases we are faced with the problem of determining not only the
magnitude of the cut--off, but also its
nature and origin. It is difficult to take seriously statements
claiming that the cut--off must be of the order of the QCD scale, the
chiral symmetry or the confinement scale; such easy statements
fail to tell us how precisely to regulate the quark loop, or what the
physical mechanism is which  provides the cut--off. For some observables
(meson mass ratios, soliton energies, the effective potential, etc.)
it does not seem to matter very much which regularization is used
\ref\rjmr{M.~Jaminon, R.~Mendez-Galain and G.~Ripka, \NP\vyp{A53}{1992}
{48}.}. But many features, such as the rates
of anomalous decays,
can and indeed do depend crucially on the chosen regularization method.
\Fig{1}{The quark loop.}{2in}

Opinions differ as to which regularization is to be used.\foot{
A review of the definition and of various regularizations of
chiral fermion determinants, and the structure of the resulting chiral
anomalies, may be found in
ref.\ref\rrdbpr{R.~D.~Ball, \PRep\vyp{182}{1989}{1}.}.}
Nambu--Jona-Lasinio models,
for example, have been regularized by a variety of methods: 3-- or
4--momentum cut--off functions, proper--time cut--offs, Pauli--Villars
regulators, etc.
In fact proper--time regularization is useless if the fermions are
chiral since in general only the real part of the determinant is then
well defined. Pauli--Villars regularization, although leading to a
well defined determinant, necessarily breaks both global
and local axial symmetries; this then leads to all sorts of
contradictions with current algebra. For example the rates for the
anomalous processes $2K\to 3\pi$ and
$\gamma\to 3\pi$ will be given incorrectly\ref\rsrm{N.~N.~Scoccola, M.~Rho
and D.~P.~Min \NP\vyp{A489}{1988}{612}\semi
N.~N.~Scoccola, D.~P.~Min, H.~Nadeau and M.~Rho,
\NP\vyp{A505}{1989}{497}\semi
M.~Wakamatsu \AP\vyp{193}{1989}{287}.}; nonanomalous processes
will be similarly corrupted. However in strictly local chiral models no
consistent regularization which preserves global chiral symmetry is
known, apart
from the rather artificial ``LR--regularization'' in which the Dirac
spinors are split into their Weyl components, and all local symmetries
are violated. Furthermore it has been
known for some time \ref\rkrs{\"O.~Kaymakcalan, S.~Rajeev and
J.~Schechter, \PR\vyp{D31}{1985}{1109}.} that with such
regularizations even the amplitude for $\pi^0\to 2\gamma$, calculated
from the famous `triangle' diagram (see fig. 2)
contradicts the rather general (and of course phenomenologically
remarkably successful) result of Adler, Bell and Jackiw \ref\rabj{S.~Adler,
\PR\vyp{177}{1969}{2426}\semi
J.~S.~Bell and R.~Jackiw, \NC\vyp{LXA}{1969}{47}\semi
W.~Bardeen, \PR\vyp{184}{1969}{1848}.}.
\Fig{2}{The triangle diagram for $\pi^0\to 2\gamma$.}{1.5in}

Momentum cut--offs can preserve
global symmetries at the expense of non--locality.
However even so there may be problems with anomalous decays.
For example
five years ago a paper appeared \ref\rbhs{A.H.~Blin, B.~Hiller and
M.~Schaden, \ZP\vyp{A331}{1988}{75}} in which the authors calculated,
among other things, the $\pi^0\to 2\gamma$ decay rate, using the
Nambu--Jona-Lasinio model. The paper rightly took the point of view
that if a cut--off is introduced to regularize the normal parity part of the
action, the abnormal parity part should also be calculated with the same
cut--off. However using a sharp momentum cut--off of a reasonable
magnitude they found a very substantial reduction in the contribution of
fig.~2 to the decay amplitude, the result of \rabj\ only being
recovered in the infinite cut--off limit. More recently however, a
non--local model with a smooth momentum cut--off has been shown
\ref\rhtv{B.~Holdom, J.~Terning and K.~Verbeek, \PL\vyp{B232}{1989}{351}.}
 to give the correct result for $\pi^0\to 2\gamma$ independently of the precise
form of the momentum cut--off.

Because of, or perhaps in spite of, all these well--known results,
many groups are still performing calculations with Nambu--Jona-Lasinio
or chiral quark models in which the
nonanomalous (and in many
cases diverging) normal processes (resulting from terms in  the real
part of the fermion determinant) are regularized using typically
proper--time regularization, with a finite cut--off, while the
finite anomalous processes (resulting from terms in the
imaginary part of the fermion determinant)
are either added by hand, or calculated in the infinite cut--off
limit (the problems described in \refs{\rsrm,\rkrs}\ being ignored). It
seems to us that such calculations
make little sense. Physics is an art. But to calculate according to
the rule ``wherever a divergent integral occurs, chop off the
divergent part'' is not. Clearly it must be possible to regulate the
quark loop (both real and imaginary parts) with a finite cut--off
scale in such a way that global chiral symmetry is preserved, and the
correct rates for all the anomalous processes (and in particular
$\pi^0\to 2\gamma$) are reproduced, without the need for ad hoc subtractions.

In low energy effective theories, calculated observables are
in general approximately independent of the form of the cut--off if they are
non--anomalous and involve energy scales much lower than the
cut--off scale. When the energy scale of the calculated observable
(the inverse soliton radius, the mass, etc.) is of the same order of
magnitude as the cut--off, or the observable is anomalous (as are the decay
rates mentioned above), it will in general depend strongly on the value and
nature of the cut--off; the cut--off thus becomes an intrinsic feature
of the model which attempts to account for complicated interactions
which would not otherwise be included. In spite of
this, most discussions of the regularization of quark loops in low
energy chiral theories have avoided the question of its physical
origin. This question is addressed in this article. Its natural result,
namely the regularization of the quark loop by the non--locality of
its interaction with the chiral field, is found in fact to have
precisely the right properties to completely resolve all the paradoxes
related to anomalous decays.

\newsec{Dynamically Regularized Quark Loops.}

In this section we explain how the quark loop is regulated
dynamically by the intrinsic non--locality of the
quark--meson interaction. The chiral field, which in the vacuum gives
the quarks their constituent mass through a dynamical chiral
symmetry breaking mechanism, gives rise to a quark self energy
which is not a constant but a non--local function of the quark
momentum. This mass generation
may be thought of schematically as a Schwinger--Dyson self--energy in the
form of a `summation' of rainbow diagrams, or equivalently of an insertion of a
$q$-$\bar{q}$ pair interacting through a Bethe--Salpeter ladder of gluon
exchanges (see fig.~3).
\Fig{3}{a) Rainbow graphs b) Ladder graphs.}{1.5in}
\Fig{4}{The non--local coupling of
                      the chiral field to $q$-$\bar{q}$.}{1in}

Writing the quark
propagator as $S^{-1}(p)=(\pslash +\Sigma(p^2))^{-1}$, where
$\Sigma (p^2)$ is the quark self energy, the most general effective
quark--meson coupling is, by definition, the amputated Bethe--Salpeter
amplitude  (see fig.~4 for the notation)
\eqn\bsdef
{\chi (k,k')\equiv S^{-1}(k)
\matele{0}{\psi(k)\bar{\psi}(k')}{\pi(k-k')}S^{-1}(k'),}
where $|\pi(k-k')\rangle $ is a pion state of momentum $k-k'$, containing
a $q$-$\bar{q}$ pair with momenta $k$ and $k'$ respectively.
The quark--pion dynamics may then
be summarized by writing an effective action
\eqn\seff
{S_{\rm eff}=\int\!{d^4x}\;\bar\psi(x) i\slash{D}\psi(x) +
\int\!{d^4x}\int\!{d^4x'}\;\bar\psi(x)M(x,x')\psi(x'),}
where the covariant derivative
$D_\mu\psi\equiv(\partial_\mu+A_\mu)\psi$ couples the
quarks locally to external gauge fields (such as the photon), while
$M(x,x')$ is the double Fourier transform of
\eqn\emdef
{M(k,k')=f_\pi\chi(k,k')U(k-k').}
Here $U(k)$ is the Fourier transform of the chiral field
$U(x)\equiv\exp\big(i\gamma_5\pi(x)/f_\pi\big)$ ($f_\pi$ being
the pion decay constant).
Expanding $U$ in powers of $\pi$ gives the quark self energy and
then couplings of the $q$-$\bar{q}$ pair to increasing numbers of
pions $\pi$. In the chiral limit (i.e. ignoring quark masses;
we will do this throughout for simplicity)
the quark self--energy and the on--shell coupling to pions are thus related
by the chiral Ward--Takahashi identity (see for example
\ref\jj{R.~Jackiw and K.~Johnson,\PR\vyp{D8}{1973}{2386}.})
\eqn\ewti
{\Sigma(p)=f_\pi\chi (p,p).}

Effective actions such as eqn.\seff\ have long been
used to discuss dynamical chiral symmetry breaking and to compute meson
form factors and decay amplitudes
\nref\chls{C.~H.~Llewellyn~Smith {\it Ann. Phys. (NY)}
\vyp{53}{1969}{521}.}\nref\pags{H.~Pagels,
\PR\vyp{D15}{1977}{2991}; \vyp{D19}{1979}{3080};
\vyp{D21}{1980}{2337}.}\nref\pagsto{H.~Pagels and S.~Stokar,
\PR\vyp{D20}{1979}{2947}.}\refs{\chls{--}\pagsto};
more recently attempts have been made to find a formal justification
for them within QCD (\ref\msb{R.~D.~Ball, {\it Int. Jour. Mod.
Phys.}\vyp{A5}{1990}{4391}.} and references therein) and to develop their
low energy phenomenological implications more systematically
\ref\hol{B.~Holdom, \PR\vyp{D45}{1992}{2535}\semi R.~D.~Ball and
G.~Ripka, in preparation.}. In fact the action \seff\ is formally
exact in the low energy limit, in the sense that interactions of pions
mediated purely by gluons must be symmetric under a local chiral
symmetry, and are therefore trivial since
$U^\dagger U=1$\ref\rSimic{P.~Simic, \PRL\vyp{55}{1985}{40};
\PR\vyp{D34}{1986}{1903}.}.

The quark self--energy $\Sigma(p^2)$ may in principle be determined by
solution of the Schwinger--Dyson equation. In practice uncertainties
in the infrared behaviour of the gluon propagator and gluon--quark
vertex function make such a determination inpractical, except in the
deep Euclidean region. From asymptotic freedom, we expect
$\Sigma(p^2)$ to be a decreasing function for large space--like $p^2$.
Indeed it is not difficult to
show using either the operator product expansion \ref\ropeasymp{
T.~Appelquist and E.~Poggio,
\PR\vyp{D10}{1974}{3280}\semi H.~D.~Politzer, \NP\vyp{B117}{1976}{397}.}
or (less trivially) the Schwinger--Dyson equation \ref\rsdeasymp{
K.~Lane, \PR\vyp{D10}{1974}{2605}\semi R.~D.~Ball and J.~Tigg,
in preparation.} that at large Euclidean momenta
\eqn\esigmaasymp{\Sigma(p^2)\p2inf\frac{(\log p^2)^{d-1}}{p^2},}
where $d$ is a related to the anomalous dimension of $\bar\psi\psi$.
In the infrared $\Sigma(p^2)$ remains finite, but is otherwise unknown.
The scale of $\Sigma (p^2)$ is again in principle given by
solution of the (nonlinear) Schwinger--Dyson equation in terms of
$\Lambda _{QCD}$, but
in practice it is better to use the Pagels--Stokar condition\pagsto\
which fits $\Sigma (p^2)$ to the pion decay constant $f_\pi=93$~MeV:
\eqn\effnorm
{f_{\pi}^2={N_c\over 4\pi^2}\int_0^{\infty}\!dp^2\,
p^2 \Sigma (p^2) \frac{\Sigma (p^2)-\half p^2
\frac{d \Sigma (p^2)}{dp^2}}{\left(p^2+\Sigma (p^2)^2\right)^2}.}

The amputated Bethe--Salpeter amplitude\bsdef\ is rather more
complicated. In what follows we will for simplicity approximate
it by simply enforcing the identification
\ewti\ even when $k'\neq k$. In this way all the various derivative
couplings of pions to quarks are suppressed, and the non--derivative
coupling is given (in the chiral limit) simply by the quark self
energy. We will explain elsewhere why, for the
processes considered here, this approximation is in practice a very
good one.

The soft asymptotic behaviour \esigmaasymp\ of the dynamically induced
quark self--energy, and thus of the effective quark--pion coupling,
means that if we use the effective action \seff\ to calculate quark
loops in which at least some of the external legs are pions, such
loops will always be finite. In this way the form of
the regularization of the quark loop (by which we mean both its
real and imaginary parts) in the effective theory of quarks and pions
is determined by the underlying theory from which it is in
principle derivable, namely QCD. If this natural regularization
is ignored (by, for example,
replacing the non--local vertex functions with local ones)
unphysical divergences are created, which must then be regulated by
some ad hoc procedure; if this procedure is not chosen properly,
the quark loop may not be well--defined, and inconsistencies of the
type described in \refs{\rsrm,\rbhs}\ can arise. Unpopular as non--local
field theories may be (although as shown in
\ref\rbt{R.~D.~Ball and R.~S.~Thorne, OUTP-93-23P, CERN-TH.7067/93}, if the
non--locality is particularly mild (i.e. analytic), many of the traditional
objections to such theories may be shown rigorously to be without
foundation), the non--locality of effective low energy
Lagrangians is in general an unescapable fact that should be faced
squarely. In the following sections we sketch one useful consequence of the
non--locality; it allows us to recover the correct (and indeed
experimentally confirmed) results for anomalous processes.

\newsec{The Regularized Fermion Determinant}

The inverse quark propagator is now described by a Dirac operator of the
form:
\def\Do{{\cal D}}
\eqn\edirac
{\Do = -i\gamma_\mu \partial_\mu + M}
where $M$ is the non--local self energy \emdef, which (using
\emdef\ and \ewti) is assumed to be of
the form (expressed for convenience in a (bra)(ket) notation, with
$\langle x|k\rangle =\Omega^{-d/2}e^{ix\cdot k}$)
\eqn\eself{
\matele{x}{M}{y}=\Sigma(x-y)U\big(\half(x+y)\big),}
where $\Sigma(x)$ is real and $U(x)U^\dagger(x)=1$. The non--locality
is thus represented by a self--energy function which is diagonal
in momentum space:
\eqn\eSigma{
\matele{q}{\Sigma}{k}=\delta_{qk}\frac{1}{\Omega^d}\int
\!d^dx\,e^{iq\cdot x}\Sigma(x)\equiv
\delta_{qk}\frac{1}{\Omega^d}\Sigma(q),}
where $\Omega^d$ is the volume element in $d$-dimensional space--time,
while the dynamical chiral field $U(x)$ is local (diagonal in
$x$-space).
%
%
A matrix element of $M$ between plane wave states can then be expanded
as follows:
\eqn\eMexp
{\eqalign{\matele{q}{M}{k}
&=\frac{1}{\Omega^d}\Sigma\big(\half(q+k)\big)U((k-q)\cr
&=\Big\langle q\Big\vert\Big(U\Sigma+\smallfrac{1}{2i}U_\mu\Sigma_\mu
+\smallfrac{1}{2!(2i)^2}U_{\mu\nu}\Sigma_{\mu\nu}+\cdots\Big)
\Big\vert k\Big\rangle,}}
where $\Sigma(k)$ and $U(k)$ are the Fourier transforms
$\Sigma(k)=\int\! d^4k\,e^{ik\cdot x}\Sigma(x)$, etc., and $\Omega^d$
is the volume element in $d$-dimensions. The second line of \eMexp\ is
written as an operator product expansion with the notation
\eqn\ederivnot
{\Sigma_{\mu\nu\cdots}(k)
=\frac{\partial\cdots}{\partial k_\mu\partial k_\nu\cdots}
\Sigma(k),\qquad
U_{\mu\nu\cdots}(k)=\int\! d^4k\,e^{ik\cdot x}
 \frac{\partial\cdots}{\partial x_\mu\partial x_\nu\cdots} U(x).}
{}From \eMexp\ we can derive an expansion of the Dirac operator \edirac\
in powers of the gradients of the chiral field $U$.

The contribution of the fermion loop to the action is formally
\eqn\eloopact
{S_D\equiv-\Tr\ln\Do\equiv S_D^+ + S_D^-.}
In practice for local couplings to external fields
$S_D$ may be defined by integrating its functional
derivatives. This is only a good definition if the regularization
of the fermion loop is chosen in such a way that this integration is
path independent\ref\rleut{H.~Leutwyler, \PL\vyp{B152}{1985}{78}.}. When
this is the case the real (normal parity) part is given by
\eqn\eactreal
{S_D^+=-\half\Tr\ln{\Do^\dagger\Do}\equiv\half\int_0^\infty
\frac{d\tau}{\tau}\Tr\,\matele{x}{e^{-\tau\Do^\dagger\Do}}{x}}
in just the same way as for a bosonic field, since
\eqn\eddaggerd
{\Do^\dagger\Do =-\partial^2-\gamma_\mu[\partial_\mu,M]+M^\dagger M}
is both hermitian and positive definite. The second two terms may be
expanded in derivatives of $U$ in an analogous way to \eMexp, and thus
a derivative expansion for the real part of the action obtained from
\eactreal\ in much the usual way\rrdbpr. The leading term in this
expansion (the `Weinberg Lagrangian') then yields precisely the
expression \effnorm\ for the pion decay constant; the terms with four
derivatives give the Gasser--Leutwyler coefficients \hol.

An exact representation for the imaginary (or abnormal parity)
part may also be constructed\nref\rboag{R.~D.~Ball and H.~Osborn,
\PL\vyp{B165}{1985}{410}, \NP\vyp{B263}{1986}{245}\semi
R.~D.~Ball \PL\vyp{B171}{1986}{435}\semi
L.~Alvarez--Gaum\'e, S.~Della--Pietra and V.~Della--Pietra,
\PL\vyp{B166}{1986}{177}.}
\refs{\rboag,\rrdbpr}:
\def\Q6{{\bf Q}}
\def\j6{{\bf j}}
\def\Man#1{{\cal M}^{#1}}
\eqn\eactimag
{S_D^-=2\pi i\Q6 = 2\pi i\int_{\Man{d+1}}\!d^{d+1}x\,\j6_0^{V-},}
where $\Man{d+1}$ is an open manifold in $d+1$ dimensions whose
boundary is the space--time $d$--dimensional manifold $\Man{d}$, and
$\j6_\mu^{V-}$ is the abnormal parity part of the conserved vector
current in $d+2$ dimensions; $\Q6$ is then the conserved charge on the
open manifold $\Man{d+2}$. It is crucial that the current
$\j6_\mu^{V-}$ is conserved, since this guarantees that the action
$S_D$ is independent, mod $2\pi i$, of the extension of $\Man{d}$
to $\Man{d+1}$\ref\rwitt{E.~Witten, \NP\vyp{B223}{1983}{422}.}. This
in turn is closely related to the integrability condition\rleut; when
there is an integrability obstruction, the conservation of the current
will be violated anomalously. The derivative expansion of the
representation \eactimag\ may be constructed by expanding the current;
the leading term is then the famous Wess--Zumino term, while all
subsequent terms (those with at least $d+3$ derivatives) may be shown
to be total derivatives, and thus local.

A similar construction may be carried out when some of the couplings of the
bosonic fields to the quark loop are non--local, as in \seff\ and \edirac. It
turns out that the regularization due to the form--factor satisfies
the integrability condition, while naturally preserving the global
$U(n_f)\times U(n_f)$ chiral symmetries (though of course local chiral
symmetries may still be broken anomalously). The representations
\eactreal\ and \eactimag\ remain the same, although now more care must
be taken to construct a conserved vector current; this will be
considered in more detail in the following section.

\newsec{The Vector Singlet Current}

As explained in the previous section the calculation of the abnormal
parity part of the quark effective action can be
achieved by first calculating the abnormal parity part of the vector
singlet current as an expansion in derivatives. The full calculation
will be explained in a forthcoming publication. We give here a
simplified account
which focusses on the relevant features of the non--locality. We
neglect here contributions from the current quark masses, and
set the external gauge fields to zero; we will explain how to extend
our results to the gauged current in the next section.

We construct the vector current using the Noether construction. When
the Dirac operator undergoes a gauge transformation,
\eqn\egt
{\Do\to e^{-i\alpha(x)}\Do\,e^{i\alpha(x)},}
the first order variation of the action is
\eqn\edeltaact
{\eqalign{\delta S_D&=i\Tr(\Do\inv[\alpha,\Do])\cr
&= -\Tr\big(\Do\inv(\alpha_\mu\gamma_\mu -i[\alpha,M])\big)
\equiv\int\!d^dx\,\alpha_\mu(x)j_\mu^V(x),\cr}}
where $j_\mu^V(x)$ is the vector current.

In local theories the commutator $[\alpha,M]$ appearing in \edeltaact\
vanishes and we recover the familiar contribution from the operator
$\bar{q}\gamma_\mu q$. The commutator $[\alpha,M]$ thus gives contributions
to the current which have been induced by the nonlocality; indeed this
extra term makes an essential contribution, since
without it the current would not be conserved.
It can be expanded in gradients of the chiral
field $U$; in the notation of eqn.\eMexp\ we obtain
\eqn\ecomexp
{[\alpha,M]=\alpha_\mu (iU\Sigma_\mu
+\half U_\alpha\Sigma_{\alpha\mu}+\cdots)
+\half\alpha_{\mu\nu}(U\Sigma_{\mu\nu}-\cdots)+\cdots.}
We thus deduce the following expression for the vector singlet
current;
\eqn\eveccurrexp
{j_\mu^V(x)=-\Tr\matele{x}{(\gamma_\mu+U\Sigma_\mu
-\smallfrac{i}{2}U_\nu\Sigma_{\nu\mu}+\cdots)\Do\inv}{x}.}

The current may be readily separated into real and imaginary parts:
\eqn\erealimag
{j_\mu^V(x)=\half(j_\mu^V+(j_\mu^V)^*)+\half(j_\mu^V-(j_\mu^V)^*)
\equiv j_\mu^{V+}+j_\mu^{V-}.}
Here we will consider only the imaginary part $j_\mu^{V-}$ which
describes abnormal parity processes. Noting that $\Do\inv=\Do^\dagger
(\Do\Do^\dagger)\inv$, and using \eveccurrexp, we may write
\eqn\eapvcexp
{j_\mu^{V-}=\Tr\int_0^\infty\! d\tau\,
\matele{x}{(\gamma_\mu+U\Sigma_\mu
-\smallfrac{i}{2}U_\nu\Sigma_{\nu\mu}+\cdots)
(\Do^\dagger e^{-\tau\Do\Do^\dagger} + \Do e^{-\tau\Do^\dagger\Do})}{x}.}

The expressions \eMexp\ and \eapvcexp\ can then be used to derive
an expansion of the abnormal parity vector current in gradients of
the chiral field $U$.
Collecting all the terms we find that, to lowest
order in the gradient of the chiral field $U$, the current is given by
\eqn\etopcurr
{j_\mu^{V-}=i^{d/2}N_c\epsilon_{\mu\nu_1\nu_2\cdots\nu_{d-1}}
\tr\big(\bar\gamma
U^\dagger U_{\nu_1}U_{\nu_2}^\dagger\cdots U_{\nu_{d-1}}\big)
\frac{1}{(2\pi)^d}\int\!d^dk\,
\frac{\big(\Sigma^d-2\Sigma^{d-1}k^2\frac{d\Sigma}{dk^2}\big)}
{(k^2+\Sigma(k)^2)^d}+\cdots.}
Now if we set $y=k^2$ and $\Sigma(k)^2=y\phi(y)$, the integral in
\etopcurr\ reduces to
\eqn\etotderiv
{\eqalign{\frac{\pi^{d/2}}{\Gamma(\smallfrac{d}{2})}
\int_0^\infty\,(k^2)^{\smallfrac{d}{2}-1}dk^2\,
\frac{\big(\Sigma^d-2\Sigma^{d-1}k^2\frac{d\Sigma}{dk^2}\big)}
{(k^2+\Sigma(k)^2)^d}
&=-\frac{\pi^{d/2}}{\Gamma(\smallfrac{d}{2})}\int_0^\infty\!dy\,
\frac{\phi^{\smallfrac{d}{2}-1}\phi'}{(1+\phi)^d}\cr
&=\frac{\pi^{d/2}}{\Gamma(\smallfrac{d}{2})}\int_0^\infty\!d\phi\,
\frac{\phi^{\smallfrac{d}{2}-1}}{(1+\phi)^d},\cr}}
where in the second line we use the asymptotic behaviour \esigmaasymp\
to show that \hbox{$\phi(y)\sim y^{-3}\to 0$} as $y\to\infty$, while we
assumed (with little loss of generality) that as $y\to 0$, $\Sigma(y)$
is strictly bounded below by $\sqrt y$, so $\phi(y)\to\infty$.
{}From \etotderiv\ it may be seen that the integrand was an exact
differential; the coefficient of the topological current \etopcurr\ is thus
independent of the precise form of $\Sigma(k)$, and is in fact the same
as in the local model in which $\Sigma(k)=m$. Note that, had
we omitted the extra contribution $[\alpha,M]$ to the current
\edeltaact, which arises from the non--locality of the mass operator
$\Sigma$, the integrand in \etopcurr\ would not have been an exact
differential and we would not have obtained a contribution which is
independent of $\Sigma(k)$.

In fact the integral is just one of the representations of the
standard function
\hbox{$\beta(z,w)\equiv\Gamma(z)\Gamma(w)/\Gamma(z+w)$}.
%
%
%
The topological current \etopcurr\ thus takes the usual form
\nref\rzee{R.~Aviv and A.~Zee, \PR\vyp{D5}{1972}{2372}}
\nref\rgw{
L.~Faddeev, {\it Lett. Math. Phys.~}\vyp{1}{1976}{289}\semi
J.~Goldstone and F.~Wilczek, \PRL\vyp{47}{1981}{986}.}\refs{\rzee,\rgw}
\eqn\egwcurr
{j_\mu^{V-}=\frac{i^{d/2}(\smallfrac{d}{2}-1)!}{(4\pi)^{d/2}(d-1)!}
\epsilon_{\mu\nu_1\nu_2\cdots\nu_{d-1}}
\tr\big(\bar\gamma
U^\dagger U_{\nu_1}U_{\nu_2}^\dagger\cdots U_{\nu_{d-1}}\big).}
Of course this is as it must be; if it were not the winding number of
the chiral field $U$ would not be equal to its topological charge.
Substitution of \egwcurr\ into \eactimag\ successfully reproduces
Witten's form \rwitt\ of the Wess--Zumino term, and thus the standard
result for the anomalous process $2K\to 3\pi$.

\newsec{Gauged Currents and the Wess--Zumino term}

Since the gauged action is required when considering the processes
involving external photons, and thus for the amplitudes of such
processes such as $\pi^0\to 2\gamma$ and $\gamma\to 3\pi$, it is
necessary to extend the calculation described in the previous section
to include local couplings to external gauge fields. Fortunately this
may be done without too much difficulty.

We thus consider a more general Dirac operator of the form
\eqn\ediracgauged
{\Do = -i\gamma_\mu D_\mu + M;}
\eddaggerd\ thus becomes
\eqn\eddaggerdgauged
{\Do^\dagger\Do =-D^2+M^\dagger M-\gamma_\mu [D_\mu,M]
-\sigma_{\mu\nu}F_{\mu\nu}.}
These simple expressions belie the full complexity of the gauge field
dependence, however, since as $M$ is now non--local, it must also
depend on the gauge field in order to maintain gauge invariance. Expansions
such as \eMexp\ will thus become rather more complicated. Although
this must be taken into account when expanding the normal parity part
of the action \eactreal\ \hol, it is fortunately of no consequence for
the leading term in the expansion of the imaginary part \eactimag\
since there each derivative of $U$ must be contracted with a
gamma--matrix if it is to lead to a non--vanishing contribution. Thus
for present purposes we may ignore the gauge field dependence of $M$.

A further complication is that for general vector and axial gauge
fields \hbox{$A_\mu=v_\mu+\bar\gamma a_\mu$} the integrability obstruction
\rleut\ will no longer always vanish. The representations \eactreal\ and
\eactimag\
must then be supplemented by some consistent regularization scheme
(for example Pauli--Villars) just as in ref.\rboag, resulting in the
usual external field anomalies. It is not difficult to check that
these anomalies depend only on the external fields; at the regulator
scale the effects of the soft couplings to the pions (and indeed to
any other strongly bound states) are effectively
screened out by their form factors. In the standard model the
anomalies are of course cancelled off by similar contributions
from the leptons.

It follows that in the presence of the external gauge fields the
gauged current corresponding to \etopcurr\ is obtained by letting
$U_\mu=D_\mu U\equiv\partial_\mu+\hat{A}_\mu U-UA_\mu$ (where
$\hat{A}_\mu\equiv v_\mu-\bar\gamma a_\mu$), adding the extra
contributions due to the last term in \eddaggerdgauged, and then
finally (to
ensure that the current is conserved and the determinant well defined)
subtracting off the anomalous terms which depend only on the external
fields. The only nontrivial step is thus the second, and the new
contributions may be found by replacing either
$U_{\nu_p}U_{\nu_{p+1}}^\dagger$ by $F_{\nu_p\nu_{p+1}}$ or
$U_{\nu_p}^\dagger U_{\nu_{p+1}}$ by
$U^\dagger\hat{F}_{\nu_p\nu_{p+1}}U$ in all possible ways. Since the
couplings to the external fields are local, the number of factors of
$\Sigma$ under the integral is reduced by two for each replacement,
while the number of factors of $k^2+\Sigma^2$ is reduced by one; for
a term with $f$ factors of $F_{\mu\nu}$ the integral is thus
$\beta(\smallfrac{d}{2}-f,\smallfrac{d}{2}$), again completely
independent of the form of $\Sigma(k^2)$. Indeed using the symmetrical
representation for $\beta(m,n)$
%
%
the gauged current may be seen to take the form
\eqn\egaugedcurr
{\eqalign{\frac{i^{d/2}N_c}{(4\pi)^{d/2}\Gamma(\smallfrac{d}{2})}
\epsilon_{\mu\nu_1\nu_2\cdots\nu_{d-1}}&\half\int_{-1}^{1}\!dt\,
\tr\bar\gamma\Big[U^\dagger U_{\nu_1}
\prod_{p=1}^{\smallfrac{d}{2}-1}\Big(
\half(1+t)U^\dagger\hat{F}_{\nu_{2p}\nu_{2p+1}}U\cr
&+\half(1-t)F_{\nu_{2p}\nu_{2p+1}}
+\half(1+t)\half(1-t)U_{\nu_{2p}}^\dagger U_{\nu_{2p+1}}\Big)\Big].\cr}}
Not surprisingly this is just the standard expression for the
Chern--Simons form $\omega^{VA}_{d-1}$ (see for example eqn.(3.100) of
ref.\rrdbpr). We thus recover the usual form for the gauged
topological current in arbitary even dimension $d$.

It now follows immediately from the representation \eactimag\ that the
leading term in the derivative expansion of the abnormal parity part
of the effective action is indeed given by the standard Wess--Zumino
term in (using Pauli--Villars to regulate the external currents) VA
form. In four--dimensional space--time (which means we should take
$d=6$ in \egaugedcurr) this means in particular that when the pion
couplings to the quark loop are regulated naturally by their soft form
factors, the amplitudes for the processes $\pi^0\to 2\gamma$ and
$\gamma\to 3\pi$ (and indeed that for $KK\to 3\pi$ in $SU(3)$) are
given to leading order in the derivative expansion and chiral
perturbation theory by the results obtained in the local chiral quark
model in the infinite cut--off limit.

The paradox noted in ref.\rbhs\ is resolved by noting that the
sharp momentum cutoff used there is equivalent to taking
$\Sigma(k^2)=m\Theta(\Lambda^2-k^2)$. Then $d\Sigma/dk^2
=-m\delta(k^2-\Lambda^2)$, and the second term in \etopcurr\ results
in a positive boundary contribution to the integral which vanishes
as $(m/\Lambda)^d$ in the limit $\Lambda/m\to\infty$. When this
contribution (which results from the extra commutator term in the
current \edeltaact, and is necessary for its conservation) is ignored,
the remaining piece is then too small to account for the full
amplitude.

\newsec{Conclusion}

The non--locality of the quark mass term $\Sigma$, and thus of the
quark--meson interactions, provides us
with a consistent and physically motivated regularization of the quark
loop in which the real and imaginary parts of the action are treated
on the same footing and in which the leading term in the derivative
expansion of the abnormal parity part of the action becomes
independent of $\Sigma$ so that we recover the usual results
\refs{\rabj,\rzee}\ for the anomalous decays $\pi^0\to 2\gamma$ and
$\gamma\to 3\pi$. Indeed, apart from mass corrections (due both to
contributions from the current mass terms in the quark loop,
and from higher order terms in the derivative expansion) and pion loop
corrections (which only begin at next-to-leading order in the
derivative expansion, at least at one loop\ref\rdw{
J.~F.~Donoghue and D.~Wyler \NP\vyp{B316}{1989}{289}.}) the naive
calculation with a pointlike pion and bare quark loop gives the same
result as the highly non--perturbative effective theory. The
Adler--Bardeen theorem\ref\rab{S.~L.~Adler and W.~A.~Bardeen,
\PR\vyp{182}{1969}{1517}.} may thus be confirmed not only to all orders
in perturbation theory, but non--perturbatively as well.

\bigskip
\noindent{\bf Acknowledgement:} RDB would like to thank the Royal
Society for their financial support, and Mannque Rho for bringing this
problem to his attention some years ago. GR would like to thank the
Paris--Oxford programme for funding visits to the University of Oxford, where
this work was initiated.

\vfill\eject
\listrefs\vfill\eject
\listfigs\vfill\eject
\bye